\documentclass[aps,reprint,superscriptaddress,noeprint,amsmath,amssymb,prb,showkeys,floatfix,nofootinbib]{revtex4-2}

\usepackage{amsmath}
\usepackage{amssymb}
\usepackage{calrsfs}
\usepackage[mathscr]{euscript}
\usepackage[version=4]{mhchem}
\usepackage{makecell}
\usepackage[caption=false]{subfig}
\usepackage[dvipsnames]{xcolor}
\usepackage{graphicx}
\usepackage[hidelinks=true,colorlinks=true,linkcolor=blue,citecolor=blue]{hyperref}
\usepackage[utf8]{inputenc}
\usepackage[T1]{fontenc}
\usepackage{physics}
\usepackage{mathtools}
\usepackage{siunitx}
\usepackage{bm}
\usepackage{dcolumn}
\usepackage{appendix}
\usepackage{lipsum}
\usepackage{multirow}
\usepackage{placeins}
\usepackage{verbatim}
\newcommand{\orcid}[1]{\href{https://orcid.org/#1}{\includegraphics[width=8pt]{orcid.pdf}}}
\usepackage[normalem]{ulem}

\let\oldAA\AA
\renewcommand{\AA}{\text{\normalfont\oldAA}}

\newcolumntype{P}[1]{>{\centering\arraybackslash}p{#1}}

\begin{document}

\title{Machine-Learning-Assisted Construction of Ternary Convex Hull Diagrams}
\author{H. Rossignol}
\affiliation{School of Physics and CRANN Institute, Trinity College, Dublin 2, Ireland}
\author{M. Minotakis}
\affiliation{School of Physics and CRANN Institute, Trinity College, Dublin 2, Ireland}
\author{M. Cobelli}
\affiliation{School of Physics and CRANN Institute, Trinity College, Dublin 2, Ireland}
\author{S. Sanvito}
\affiliation{School of Physics and CRANN Institute, Trinity College, Dublin 2, Ireland}
\date{\today}

\begin{abstract}
In the search for novel intermetallic ternary alloys, much of the effort goes 
into performing a large number of \textit{ab-initio} calculations covering a 
wide range of compositions and structures. These are essential to build a reliable convex
hull diagram. While density functional theory (DFT) provides accurate predictions 
for many systems, its computational overheads set a throughput limit on the number 
of hypothetical phases that can be probed. Here, we demonstrate how an ensemble 
of machine-learning spectral neighbor-analysis potentials (SNAPs) can be integrated
into a workflow for the construction of accurate ternary convex hull diagrams, 
highlighting regions fertile for materials discovery. Our workflow relies on 
using available binary-alloy data both to train the SNAP models and to create 
prototypes for ternary phases. From the prototype structures, all unique ternary
decorations are created and used to form a pool of candidate compounds. The SNAPs 
are then used to pre-relax the structures and screen the most favourable prototypes,
before using DFT to build the final phase diagram. As constructed, the proposed 
workflow relies on no extra first-principles data to train the machine-learning 
surrogate model and yields a DFT-level accurate convex hull. We demonstrate its 
efficacy by investigating the Cu-Ag-Au and Mo-Ta-W ternary systems.

\end{abstract}


\maketitle

\section{Introduction}


Systematic materials design aims to develop methods that can help to accelerate 
the discovery of compounds with tailor-made properties, fit for certain applications.
The large investment in the area, not least through the materials genome 
initiative~\cite{MGI}, underpins the importance of searching for novel compounds to
bolster technological progress. Atomistic simulations provide a suitable pathway to
achieve this goal, since the search can be performed systematically, at low cost and
with a complete control over structure and composition. Density functional theory (DFT)
calculations are notably used to predict material properties \textit{in silico}, such 
as material stability or elastic responses. By performing property predictions across 
a large range of prototype structures, in the form of high-throughput 
studies~\cite{NatureMat}, novel magnetic \cite{sanvito2017accelerated}, 
high-hardness \cite{coreyhea} and battery materials \cite{kim2014analysis} have 
been discovered. Extensive databases, grouping large numbers of such calculations, 
have been created and are open to the community. These include AFLOWlib \cite{AFLOW}, 
Materials Project \cite{jain2013commentary}, OQMD \cite{OQMD} and NOMAD \cite{NOMAD}. 
While such studies remain faster than experimental investigations, the composition 
and structural spaces to be searched are incredibly large, limiting the scope of 
application of pure DFT workflows. Importantly, such limitation in sampling capacity 
becomes increasingly critical as the number of elements per compound grows, despite the 
anticipation that a majority of newly discovered compounds being highly multi-elemental 
\cite{Cormac2019}. In order to address this issue and to harness the data available 
from existing \textit{ab-initio} calculations, machine learning (ML) has proved to be 
a very powerful tool, as it typically comes at a fraction of the DFT computational cost. 

The first step in high-throughput computational studies consists in identifying 
stable compounds by finding a stoichiometry and an associated structure that can 
be formed. In order to assess the stability of a given structure, the appropriate 
convex hull diagram needs to be calculated. The proximity between a compound's 
enthalpy of formation, $\Delta H_\mathrm{f}$, and the closest tie-line on the convex hull
serves as a criterion for evaluating its stability. Lower values indicate a higher
likelihood of stability. Threshold values, typically up to $\sim$100 meV/atom, are 
used as stability cut-offs \cite{kim2018machine}. In order to speed up electronic
structure methods such as DFT, one possibility is to predict this quantity directly 
by using ML models, where compounds' compositional and structural information is 
encoded and mapped directly onto $\Delta H_\mathrm{f}$. This is otherwise known as composition
prediction, as it is used to identify which stoichiometries are stable by fixing 
structural variations. The ML models typically used include neural networks, kernel 
ridge regression and random trees, while the training data are often taken from OQMD, 
AFLOWlib or Materials Project. For instance, models where the feature vector is only
based on compositional information have been used to predict the stability of compounds
forming a set prototype structures (elpasolites, perovskites, heuslers, etc.), 
which is fixed for the compounds in the training set 
\cite{faber2015crystal,faber2016machine,schmidt2017predicting,kim2018machine}. 
Including structural information in the definition of a model mainly improves the
predictions, if large training datasets ($>$10$^{5}$ data points) are 
used \cite{ward2017including}. Graph convolutional neural networks 
\cite{xie2018crystal,park2020developing,schmidt2022large} have notably been used to 
predict convex hull distances accurately and benefit greatly from structural 
features \cite{bartel2020critical,pandey2021predicting}. Note that these can also be 
constructed with compositional information only 
\cite{goodall2020predicting,jha2018elemnet}. One downside to the inclusion of 
structural information in the models is that the optimised structure is not known prior
to the search, so that data for unrelaxed structures has to be used. This can notably be 
corrected by using machine-learning interatomic potentials (MLIAPs), that are capable 
of performing relaxations.

MLIAPs combine atomic fingerprints, representing individual atomic environments in 
the form of feature vectors, with ML algorithms, and effectively map the potential
energy surface of a collection of atoms \cite{bartok2013representing}. The last decade
has seen an immense expansion of the development and application of such potentials
\cite{GAP,NNP,SNAP,MTP,ACE,JL}. When trained using active learning, MLIAPs have most
notably been able to extend the length and timescales of \textit{ab-initio} molecular 
dynamic simulations by several orders of magnitude 
\cite{deringeramorphousC,caro2020machine,jinnouchi2019fly}. 
Such potentials have been successfully applied to predict the energy and forces of 
alloys \cite{chen2017accurate}, and have been used to accelerate and assist the
construction or further exploration of binary and ternary convex hulls. Workflows built
on these potentials use MLIAPs as surrogate models to first relax and then 
make energy predictions on a large library of prototype structures. The lowest energy
structures are then compared to a reference convex hull obtained from DFT calculations. 
This process allows one to improve the reference convex hull diagram by identifying 
structures lying below it. The training of such potentials is crucial for adequate
performance, and studies insist on using high-energy structures for the relaxations to
be reliable. 

Work in this area has broadly been split into two categories. In the first, specific 
MLIAPs are trained for a given system \cite{gubaev2019accelerating,bernstein2019novo,artrith2018constructing,kharabadze2022prediction,seko2020machine,kelvinius2022graph}, typically using active learning. In the
other, MLIAPs are trained on large generic databases, and are used to scan over many
phases \cite{law2022upper,chen2022universal}. The former is more accurate 
than the latter, but it is not transferable to other phases. Due to their higher 
accuracy, phase-specific MLIAPs can also be regarded as global structure optimisers, 
in that not only they can be used to identify specific stable compositions, but they 
can accurately predict their structure as well. Many other ML global structure
optimisers, exist, either in the form of novel workflows 
\cite{bisbo2022global,paleico2020global,yamashita2018crystal,podryabinkin2019accelerating} or by inserting MLIAPs into the pre-existing state-of-the-art global structure
optimisers \cite{deringer2018data,pickard2022ephemeral,tong2018accelerating}. 

In this work, we demonstrate how a MLIAP can be trained on data {\it readily available} 
on a mainstream repository, such as AFLOWlib~\cite{AFLOW}, and used to screen a library of 
ternary-alloy prototypes constructed from their associated binary systems. 
Recently, we have shown
\cite{SNAPbi2ter} that an ensemble of spectral neighbor-analysis potentials (SNAPs) 
\cite{SNAP} models, trained on the energy data 
of the three binary subsystems associated with a ternary one, was able to predict the 
energies of ternary compounds with a mean absolute error (MAE) of $\sim$30 meV/atom, 
as long as the structures were fully relaxed. This, not only provides a fast 
energy-screening tool for ternary compounds, which only requires existing 
\textit{ab-initio} data on binary structures, but it also gives the valuable 
insight that chemical environments within binary and ternary transition-metal alloys 
are similar. Such observation is at the heart of the workflow introduced here. 
A selection of binary structures, those close to their respective convex hull
tie-plane, are selected as templates for ternary alloys. In a high-throughput set up, 
these are screened using an ensemble of SNAP models, trained on binaries. The 
lowest-energy compounds are then selected as the most promising candidates, and 
their energies are calculated using high-fidelity DFT. The ternary convex hull is 
thus updated.

What makes this workflow different to tailor-made MLIAPs used for convex hull 
construction is that all the data, both for the prototypes generation and for training
the SNAPs, are taken from the relevant binary phases of the AFLOWlib database 
\cite{AFLOW}. In other words, there is no need to generate any new data for the purpose
of training the MLIAPs. Despite its training database not being specifically made, 
either by including important configurations through physical intuition or through 
active learning, it still has a low enough error on energy predictions to enable a 
high-throughput search of novel alloys. This is because stable binary and ternary 
phases, at least for the materials class of transition-metal intermetallics 
investigated here, share similar local atomic environments. In some sense, the 
workflow enables an interpolation of the data already available in AFLOWlib, to 
scan ternary convex hulls and identify stable compositions. Since only a few high-energy
structures and no out-of-equilibrium configurations are included in the SNAP training
dataset, additional features are introduced in the workflow to increase the robustness
of the predictions. These include constraints on the SNAP-driven relaxation (constant
volume and the inclusion of a maximum number of steps) as well as using an ensemble of
models. 

In this paper, the workflow used to generate novel ternary compounds is presented. The 
methodology Section \ref{sec:Methods} details how ternary prototype structures are built
from their binary counterparts, and how binary compound data from AFLOWlib \cite{AFLOW}
is used to train an ensemble of SNAP \cite{SNAP} models. Such SNAPs are used to relax 
and screen the ternary prototypes. Then, the results section \ref{sec:Results} presents
how the workflow is used to find stable phases for the Cu-Ag-Au and Mo-Ta-W ternary
systems. The so-constructed convex hull diagrams are finally compared with their 
available AFLOWlib counterparts, and conclusions are drawn in
Section~\ref{sec:Conclusion}.

\section{Methods}
\label{sec:Methods}

\begin{figure}[!h]
    \centering
    \includegraphics[width=8.5cm]{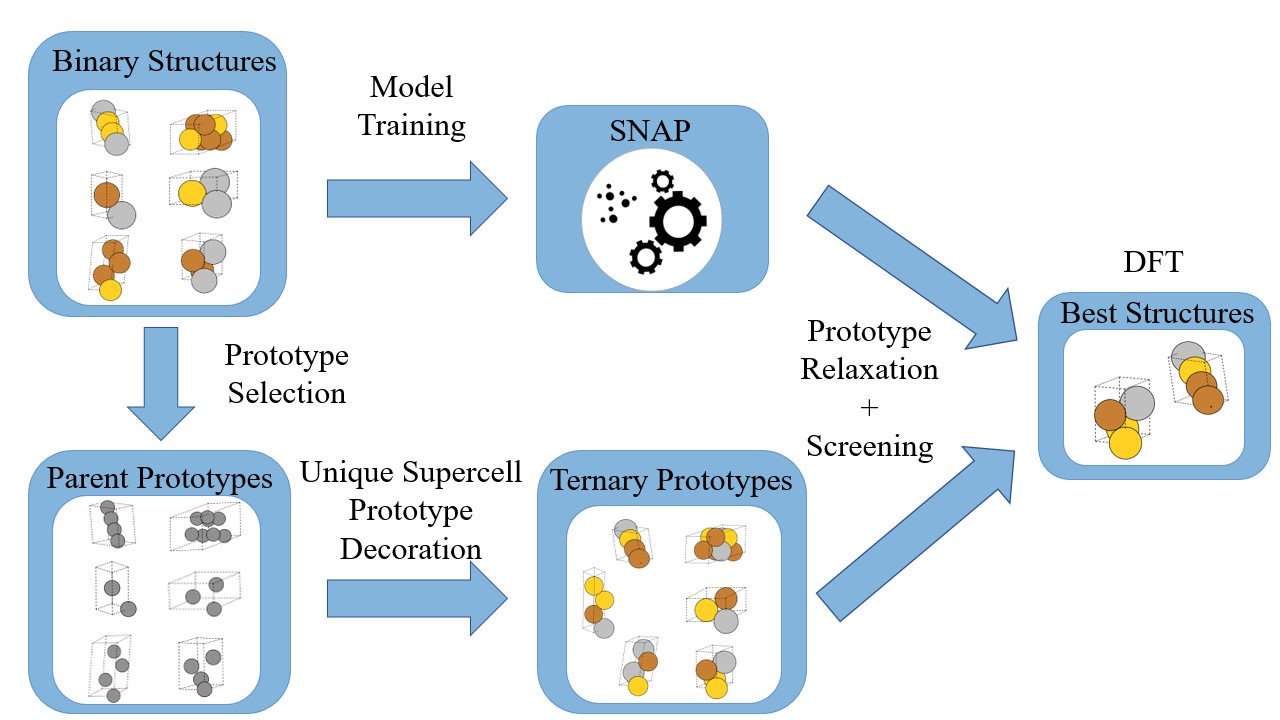}
    \caption{The stable ternary compounds search workflow implemented in this work. Data
    available from the three binary systems associated with a ternary one (box in the upper
    left corner) is used for two tasks: i) the training of an ensemble of SNAP models, and 
    ii) the construction of a library of parent prototype structures. Derivative structures 
    of the latter are created and all possible ternary decorations of these are produced. 
    Each of these are then relaxed with the SNAPs model and the lowest-enthalpy structures 
    are screened.}
    \label{fig:methodology}
\end{figure}

The general methodology of our workflow, schematically introduced in
Figure~\ref{fig:methodology}, is here described in detail. From the AFLOWlib database of 
binary compounds and their associated DFT-computed energies, an ensemble of SNAP models 
are trained. A subsection of these structures, the ones with lowest enthalpy of formation, 
are also used as parent structures and form a library of prototypes. Note that here, as is
standard practice in many DFT-based convex hull constructions, a compound's enthalpy of formation is
approximated solely by its ground-state DFT energy, so that the terms enthalpy and energy will
be considered equivalent throughout the rest of the manuscript. Candidate ternaries are then created, by generating 
all possible and unique derivative structures of such prototypes \cite{hart2008} at a fixed composition \cite{hart2012}. These 
two parts are then combined as the ensemble of SNAP models are used to relax the candidate
ternaries. The final energies are predicted from cross-validation among the ensemble of 
models, while the standard deviation of the predictions is also used to detect and remove
geometries for which the relaxation has failed. The resulting structures with lowest energy 
and standard-deviation are selected as best candidates (closest to the convex hull). Finally, 
full \textit{ab-initio} relaxation is performed for these. The ternary system Cu-Ag-Au is 
used to develop the methodology and is hence employed here as an example in each subsection.

\subsection{Generating Prototypes}
\label{subsec:Generation}

\begin{figure}[!h]
    \centering
    \includegraphics[width=8cm]{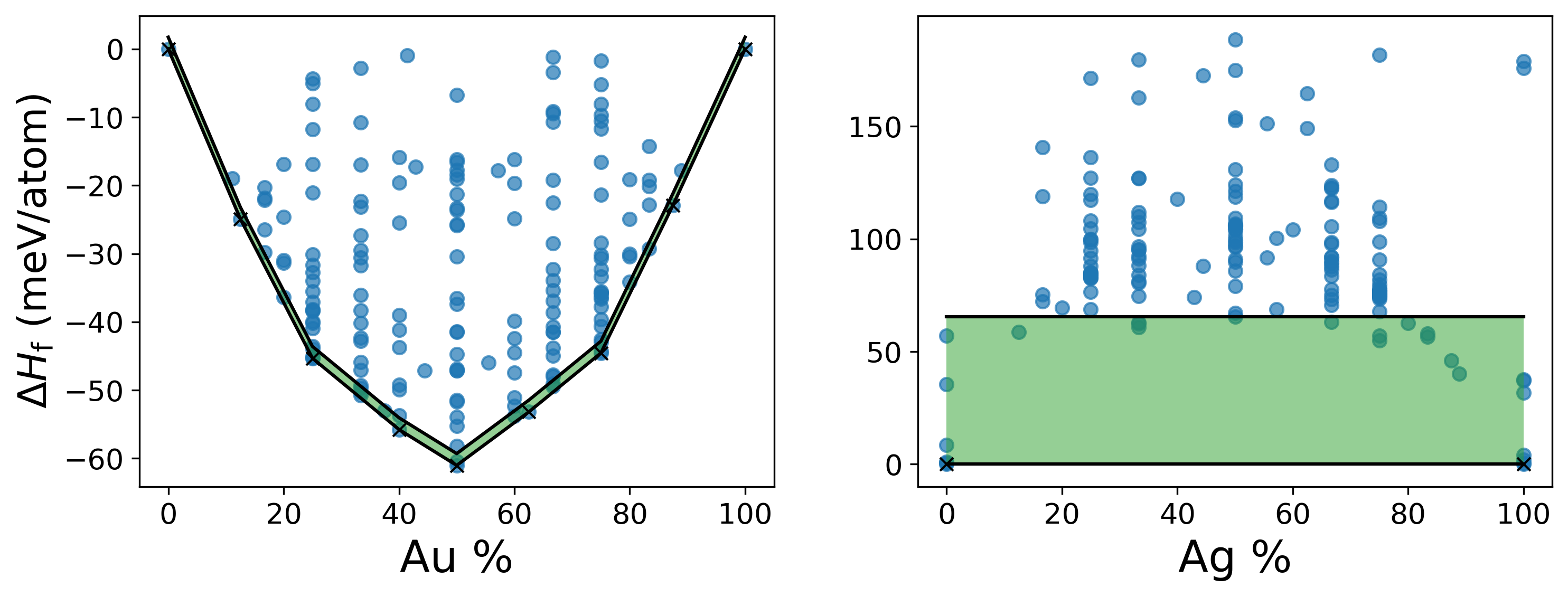}
    \caption{Convex hull diagrams of the binary Ag-Au (left) and Cu-Ag (right) binary systems. 
    The convex hull is defined by the bottom black lines, while the green shadowed regions up to the upper full lines show the energy windows chosen to select the binary structures.}
    \label{fig:CHULL_Binary}
\end{figure}

The first step of the workflow consists in creating a suitable library of ternary prototype 
materials. The driving idea of this work is that the local atomic environments seen in binary 
intermetallic alloys are similar to those in the associated ternaries, especially for
structures close to equilibrium \cite{SNAPbi2ter}. This leads to choosing the binary 
structures as prototypes for the ternaries. More specifically, the bottom of the three binary
convex hulls associated to a ternary system (in our example Ag-Au, Cu-Ag and Cu-Au for Cu-Ag-Au)
are scanned to select the lowest-enthalpy compounds. Those within a certain energy range from
the convex hull are then selected. All binary structures considered here are taken from the
AFLOWlib database \cite{AFLOW}. The threshold energy selected differs depending on the system
at hand, such as to ensure that roughly the same number of structures are taken from each
binary diagram. For instance, in our test system Cu and Ag are immiscible \cite{AgAuCuthermo}.
Therefore, all binaries have positive enthalpy of formation and lie far from the Cu-Ag tie 
line between the two elementary phases ({\it fcc} Cu and Ag). As a consequence, the energy 
window above the hull for this binary is larger than that of the other two. The convex 
hulls of Ag-Au and Cu-Ag are compared in Figure \ref{fig:CHULL_Binary} and Table 
\ref{table:BinaryStats} gives the energy window used as well as the number of structures 
selected for each binary system. 
\begin{table}[hbt]
\centering
\scalebox{1.20}{\begin{tabular}{c|c|c}\hline\hline
$X-Y$ & $N_\mathrm{struct}$ & $\Delta E$ (meV) \\
\hline
 Ag - Au  & 24 	& 1.7 \\
 Cu - Ag  & 25  	& 65.4 \\
 Cu - Au  & 25   & 6.2\\\hline\hline
\end{tabular}
}
\caption{Number of structures, $N_\mathrm{struct}$, selected from each binary system, $X-Y$, 
to construct the ternary prototypes. Here we also report the energy window, $\Delta E$, above 
the convex hull used for the selection.}\label{table:BinaryStats}
\end{table}

Once the prototypes are selected, the constituent atoms are stripped of their chemical 
identity and all structures are compared using the AFLOWlib symmetry tool \cite{hicks2018aflow}
in order to curate redundancies. This is necessary, since certain structure types (such as 
{\it fcc} or {\it bcc}) may be present several times in the collected database, but may be
``decorated'' in several different ways for different stoichiometries and binaries. All
structures are reduced to their primitive cell at this stage. It is also important to note 
that single-element structures are also included in this analysis. This leads to a library 
of unique, undecorated prototypes, taken from the binary convex hulls. For the Cu-Ag-Au 
system, this results in 40 prototypes.

From this set of prototypes, ternary alloys are generated. This task is performed at a fixed 
stoichiometry and for cells up to a maximum number of atoms, $N_\mathrm{max}$. For all the prototypes with a number of atoms compatible with the fixed stoichiometry, the set of all
unique derivative structures are created, by following the procedure introduced in 
references \cite{hart2008,hart2009,hart2012} and by using the associated open source
\textsc{Enumlib} code. The initial implementation of the algorithm begins from a parent 
lattice and uses group theory to efficiently enumerate all the unique ways to occupy the 
sites of supercells constructed from that lattice \cite{hart2008}. Further modifications of 
the scheme allow for the starting structures to be defined by a lattice and an atomic basis
(multi-lattices) \cite{hart2009} and for the generation of derivative geometries at a fixed
stoichiometry \cite{hart2012}.
This completes the first step of the workflow (bottom branch in Figure \ref{fig:methodology})
and leads to a set of unique ternary compounds, inspired by the structures of the binaries. 
The energy of these is then screened using a MLIAP.

\subsection{Ensemble of SNAP models}
\label{subsec:EnsembleSNAP}

MLIAPs typically assume that the total energy, $E$, of a $N$-atom system defined by the
coordinates $\mathbf{r}^{N}$ can be written as a sum of atomic energies $E_{i}$,
\begin{equation}
\label{eq:MLIAPE}
    E = \sum_{i=1}^{N}E_{i}\:.
\end{equation}
Such a partition, first proposed by Behler and Parinello \cite{NNP}, is based on the 
principle of near-sightedness \cite{kohn1996density,prodan2005nearsightedness}. The MLIAP 
of choice for this work is SNAP \cite{SNAP}, which has been proved to perform well regardless of
the nature of the chemical bond \cite{AleSNAP}. In this model, the total energy of a compound
is written as a sum of linear combinations of the feature vectors describing the chemical
environments of each atom $i$ of type $\alpha_{i}$ in the system. SNAP then takes the
bispectrum components, $\mathbf{B}_{i}^{\alpha_{i}}$, as feature vectors. The function, $E_\mathrm{SNAP}$, that returns the SNAP-predicted total energy is thus defined as
\begin{equation}
\label{eq:SNAPE}
    E_\mathrm{SNAP} \left ( \mathbf{r}^{N} \right ) = \sum_{i=1}^{N}\beta_{0}^{\alpha_{i}} + \boldsymbol{\beta}^{\alpha_{i}}\cdot\mathbf{B}_{i}^{\alpha_{i}}\:,
\end{equation}
where $\beta_{0}^{\alpha_{i}}$ and $\boldsymbol{\beta}^{\alpha_{i}}$ are the species-dependent
linear coefficients of the ML model. Further details on this potential can be found in 
Section~\ref{sec:CompMethods}. SNAP's linear form allows one to obtain good performance with 
a small number of features, 56 per species in our case, and when trained on small datasets 
($\leq$ 10$^{3}$ data points) \cite{MLIAPperformance}. Furthermore, the SNAP hyperparameters 
are easy to optimise, since the range of optimal values for $J_\mathrm{max}$ (the maximum angular 
momentum of the bispectrum) and $r_\mathrm{cut}$ (the cut-off radius) is wide and consistent
for accurate performance \cite{SNAP,chen2017accurate,li2018quantum}. In our experience, the 
optimisation of the atomic weights, although generally useful, only leads to modest improvements 
\cite{SNAPbi2ter}. 

As for our previous study, an ensemble of SNAP models is used to increase the robustness of 
the predictions. Furthermore, this provides a mean of estimating the prediction uncertainty 
\cite{SNAPbi2ter}. The ensemble is defined as a set of $K$ functions, 
$\left \{ E_\mathrm{SNAP}^{k} \right \}_{k=1}^{K}$, where each SNAP model,
$E_\mathrm{SNAP}^{k}$, is trained differently, and hence has different linear coefficients. 
The predicted energy of a new system with atomic positions, $\mathbf{r}^{N}$, is defined as 
the mean prediction of the models, $\bar{E}$, and its uncertainty is estimated from the
standard deviation, ${\sigma}$, of $\bar{E}$ namely,
\begin{equation}
    \bar{E}\left ( \mathbf{r}^{N} \right ) =\frac{\sum_{k=1}^{K}E_\mathrm{SNAP}^{k}\left ( \mathbf{r}^{N} \right )}{K}\:,
\end{equation}

\begin{equation}
    {\sigma}\left ( \mathbf{r}^{N} \right ) = \sqrt{\frac{\sum_{k=1}^{K}\left [ E_\mathrm{SNAP}^{k}\left ( \mathbf{r}^{N} \right ) - \bar{E}\left ( \mathbf{r}^{N} \right )\right ]^{2}}{K}}\:.
\end{equation}

The training data only consists of binary alloys obtained from the AFLOWlib database, whose
energy have been re-computed for consistency by single-point DFT calculation (no further 
relaxation is performed). Differently from what was done when generating the prototypes, here
all binaries, no matter their distance from the convex hull, are included in the training 
dataset. The same workflow is also used for Mo-Ta-W (the results are described in section
\ref{sec:Results}), for which we demonstrate that the energy values taken directly from AFLOWlib are 
suitable for training the SNAP models. The full details on the Cu-Ag-Au binary subsystems can 
be found in reference~\cite{SNAPbi2ter}.

\begin{figure}
         \includegraphics[width=0.35\textwidth]{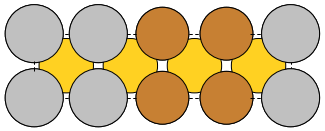}
         \\
         \includegraphics[width=0.34\textwidth]{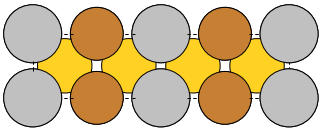}

    \medskip

    \centering
    \begin{tabular}{c||c|c}
        \hline
         \thead{$w_\mathrm{Cu}$, $w_\mathrm{Ag}$, $w_\mathrm{Au}$} & \thead{$E^\mathrm{Top}$(eV)} & \thead{$E^\mathrm{Bottom}$(eV)}\\ 
         \hline
         \hline
        1\;\;\;\;   1\;\;\;\; 1& -25.640 & -25.640 \\
        1\;\;\;\;   1\;\;\;\; 2& -25.879 & -25.879 \\
        1\;\;\;\;   2\;\;\;\;   1& -25.717 & -25.676 \\
        2\;\;\;\;   1\;\;\;\;   1& -26.162 & -26.047 \\
        3\;\;\;\;   1\;\;\;\;   2& -25.640 & -25.742 \\
         \hline
         
    \end{tabular}

        \caption{SNAP performance for two structurally identical prototypes. The upper 
        two panels show two different possible site occupations for a $3\times1\times1$ 
        {\it bcc} derivative structure with Cu$_{1}$Ag$_{1}$Au$_{2}$ stoichiometry. Here 
        we show the $z$-axis view, with brown, gray and yellow spheres representing Cu, Ag 
        and Au respectively. The table shows the SNAP-predicted energies for the two
        prototypes, when the SNAP is trained with different atomic weights, 
        $\left \{  w_{\alpha}  \right \}$, as indicated in the first column. Note that 
        when the Cu and Ag weights are identical the two energies coincide by construction.}
        \label{fig:three graphs}
\end{figure}

Previously, 10 different SNAP models within the ensemble were obtained by training on different
subsets of the same size of the binary-alloy database~\cite{SNAPbi2ter}. For this work, 5 
models are trained on the full database, but for each one, a different set of atomic weights
for Cu, Ag and Au are used to compute the bispectrum components. This difference is motivated 
by the need to distinguish compounds with identical site positions in their structure (e.g. 
the sites of a {\it bcc} supercell), but different atomic site occupations. If the atomic 
weights for all species are identical, for some of these structures, notably the high symmetry
ones, SNAP will predict identical energies for different compounds. This is illustrated in 
Figure~\ref{fig:three graphs} for two distinct compounds obtained as {\it bcc} derivative structures with 
Cu$_{1}$Ag$_{1}$Au$_{2}$ composition. Prototypes A and B only differ by a permutation of the 
Ag and Cu atoms. Hence, SNAP models using identical weights for these two atomic types will
fail to predict different energies for the compounds. Therefore, by construction, in the 
ensemble created the two elements in each pair of atomic types (e.g. Cu and Ag in Cu-Ag) have
different weights in at least one model. 

Before selecting the values for the atomic weights, $r_\mathrm{cut}$ and $J_\mathrm{max}$ are 
optimised manually and independently by using 10-fold Monte-Carlo cross-validation for 
fixed identical weights, $\{$1, 1, 1$\}$ and thus find: $r_\mathrm{cut}$~=~3.5\AA\ and
$J_\mathrm{max}$~=~4. For these values, the optimal atomic weights are set by performing a grid
search, with the same cross-validation method, and where all three atomic weights are varied 
from -5 to 5 in steps of 1. Provided the condition above is fulfilled, the sets of weights 
used for the SNAP models in the ensemble are chosen to minimise the cross-validation root-mean
squared error (RMSE). The training and cross-validation errors for each model of the ensemble
are given in Table \ref{table:SNAP_Errors}. The ensemble is then used to predict which of the
prototype structures have the lowest enthalpy.

\subsection{Energy Screening}
\label{subsec:screen}

The final aim of the workflow is to suggest low-energy ternary structures at a given 
stoichiometry. Since many compounds with a large energy spread are screened, the 
suggestions made need to be accurate (must include low-energy structures) and reliable 
(must not include high-energy and unphysical systems). While the energy error of the SNAP
surrogate model is low, it is still prone to making poor predictions on new systems that 
do not resemble the structures seen in training. As a result, the construction of the 
workflow highly focuses on the robustness of the final predictions made. Note that 
choosing parent prototypes from binary compounds already increases the reliability of 
the predictions.

\begin{table}[hbt]
\centering
\scalebox{1.10}{
\begin{tabular}{c|c|c|c|c}\hline
 \thead{$w_\mathrm{Cu}$, $w_\mathrm{Ag}$, $w_\mathrm{Au}$} & \thead{Training \\ MAE} & \thead{Training \\ RMSE } & \thead{CV \\ MAE} & \thead{CV \\ RMSE}\\
\hline
\hline
 1\;\;\;\; 1\;\;\;\; 2  & 8.0 & 13.4 & 27.1 & 83.5\\
 1\;\;\;\; 2\;\;\;\; 2  & 8.7 & 13.5 & 24.8 & 64.7\\
 -1\;\;\; -2\;\;\; -1  & 9.7 & 16.4 & 30.6 & 86.4\\
 -1\;\;\; -2\;\;\; -2 & 8.5 & 13.2 & 23.5 & 64.3\\
 -1\;\;\; -1\;\;\; -2 & 7.7 & 13.1 & 25.6 & 75.0\\
\hline
\end{tabular}
}
\caption{Training and cross-validation (CV) errors for the 5 models, defined by different 
atomic weights $\left \{  w_{\alpha}  \right \}$, of the ensemble. Here we report the mean
absolute error (MAE) and the root-mean squared error (RMSE). All values are given in meV/atom.}
\label{table:SNAP_Errors}
\end{table}

\begin{figure}
    \centering
    \includegraphics[width=8cm]{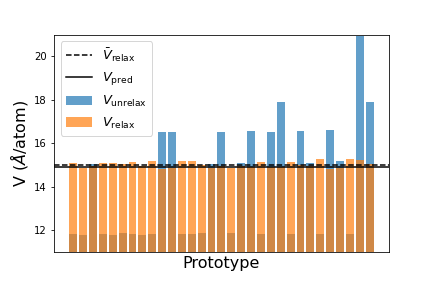}
    \caption{Plot showing the initial unrelaxed volumes, $V_\mathrm{unrelax}$, and 
    relaxed equilibrium ones, $V_\mathrm{relax}$, of a set of ternary prototypes at the
    stoichiometry Cu$_{2}$Ag$_{1}$Au$_{1}$. The unrelaxed volumes are chosen from the volume 
    of the binary associated with each ternary compounds' structure. The dashed line indicates
    the mean equilibrium volumes for these compounds, $\bar{V}_\mathrm{relax}$, while the 
    full line shows the volume predicted by the weighted average of the elemental 
    volumes, $V_\mathrm{pred}$.}
    \label{fig:volume}
\end{figure}

The first step in the energy-screening process consists in setting the compounds' unit-cell
volume. This is chosen by taking the weighted average of the elemental volumes of the 
constituent atoms, an approximation that reproduces the results from \textit{ab-initio}-relaxed 
compounds quite well, as is illustrated in Figure \ref{fig:volume}. Then, the volume and all
lattice parameters are kept fixed during any relaxation driven by the SNAP models. This is
because, while the training database includes a diverse set of structures, they are all at
equilibrium, namely their forces and stress-tensor elements are close to zero. Therefore, no
configurations are strongly compressed or expanded, a fact that causes the SNAP models to
perform poorly on the prediction of equilibrium volumes and lattice parameters. The volume is
only allowed to change for the final, most promising, structures selected for the DFT
relaxation. 

Each of the available $K$ SNAP models are used to drive an ionic relaxation, with a maximum of 
$N_\mathrm{s}$ steps, for all of the prototypes, leading to $K$ differently relaxed 
structures per prototype. A ``cross-validated'' energy prediction is given for each 
relaxed structure. Given a candidate obtained by relaxing a prototype with the $k$-th 
SNAP model, the energy prediction is made with $K$-1 models, namely all SNAPs bar the 
one used for the relaxation of the candidate at hand. The mean and standard deviation 
of the energy predictions of the $K$-1 models are then saved. For every prototype, one of 
the $K$ relaxed structures obtained is selected, namely the one with lowest ``cross-validated'' 
standard deviation. This is the structure whose final total energy has received the largest
consensus among the SNAP models. Therefore, there is only one relaxed structure per prototype.

The reason why this process is not a single ionic relaxation stems from the drive 
towards robustness of the predictions. Without the inclusion of the $N_\mathrm{s}$ iteration
cutoff, some of the relaxations would lead to structures that are trapped in unphysical 
local minima of the potential energy surface (PES) of the driving SNAP model. By stopping 
the relaxation process at a low number of steps ($N_\mathrm{s}$ = 10 in this study), this 
effect is mostly avoided, as the structures cannot change too drastically. For the relaxations 
that are accurately driven by SNAP, the largest drop in energy typically occurs during the 
first few steps of the relaxation process. While accurate relaxations are also cut before 
convergence, as they are not distinguished from the inaccurate ones, the final structures 
are still lower in energy than the initial prototypes. This reduces the likelihood of 
obtaining high-energy structures and the total run time of the workflow remains modest. 

Using ``cross-validated'' energy predictions of the relaxed structures helps to remove the 
bias of specific SNAP models. The SNAP driving the relaxation usually predicts the final 
structure to be at an energy lower than the initial one, since it moves the geometric 
configuration to, at least, a local minimum of that particular SNAP's PES. If the relaxation 
is inaccurate, the resulting structure will be, in fact, high in energy (as predicted by DFT). 
The relaxation-driving SNAP model is, therefore, not used for energy predictions. Instead, the 
other models of the ensemble are, since they are less likely to present a bias towards that relaxed structure and therefore to predict it being low in energy. 
Note that since the structures used as training data are the same for all the $K$ models, 
they could also lead to biases for the same structure. This is accounted for by using the
``cross-validated'' standard deviation, rather than the mean energy, to select the ``best''
structure out of the $K$ relaxed ones. Indeed, even if several/all SNAP models are biased
towards a particular structure and collectively predict it to have a low energy, the 
inaccurate predictions of each model, will be different. This is because they are inaccurate,
extrapolated predictions. Hence, while the mean prediction of the ensemble may give a low 
energy value, the standard deviation will be large. 

Finally the ``cross-validated'' standard deviation prediction for all structures must be 
lower than a cut-off value, $\sigma_\mathrm{cut}$, to be considered for the final energy
screening. This typically removes structures with a low SNAP-predicted energy but which DFT
predicts to be high energy, as well as structures with high SNAP-predicted energy. From the
sample of structures selected, the ones with the lowest ``cross-validated'' mean energy are
chosen and relaxed with DFT. In this study, 15 structures per stoichiometry are selected
through such process.

In summary, the workflow described creates a set of prototypes and uses an ensemble of ML
potentials to relax and screen the structures, which are most likely to have low energy. This
is done for a fixed stoichiometry. The final selected compounds are then recomputed with full
DFT relaxation. The workflow, therefore, allows one to perform all the computationally intensive
DFT calculations only on the most promising candidates. In the following section, this workflow
will be used to reconstruct the ternary-alloy convex of Cu-Ag-Au and Mo-Ta-W.

\section{Results}
\label{sec:Results}

This section, which is structured into two subsections, presents the key outcomes of 
our method. Firstly, we examine the performance of our workflow against the well-established
and extensively studied Cu-Ag-Au phase diagram~\cite{prince1990phase}. Then, we provide a
comparison between our results and those of one of the better-characterised phase diagrams
available in AFLOWlib, specifically Mo-Ta-W. By benchmarking the workflow phase diagram 
predictions to the DFT created ones available in AFLOWlib, we gain valuable insights into 
the effectiveness of our approach.

In order to accurately evaluate the stability of our predicted prototypes and ensure 
consistency in our analysis, we have used the QHull~\cite{barber1996quickhull} library to
calculate the convex hulls presented in this work. The data used to construct the reference
convex hull (a subset of the full database) was downloaded from the ground-state compounds 
calculated by AFLOWlib. In the case of the Cu-Ag-Au ternary system, in order to guarantee
consistency, we have re-calculated the energies of these compounds with the Vienna Ab 
initio Simulation Package (VASP) \cite{VASP}. Throughout the entire process, we have strictly
followed the AFLOWlib standards as outlined in reference \cite{10.1016/j.commatsci.2015.07.019}, 
with an energy cut-off of 600~eV to ensure tight convergence. More information regarding the 
DFT calculations can be found in Section \ref{subsec:DFT}. In contrast, for Mo-Ta-W we directly 
use the AFLOWlib pre-computed energies.

\subsection{Cu-Ag-Au ternary convex hull}
\label{subsec:AgAuCu}

In order to evaluate the performance of our workflow, it is essential to select a well-studied 
phase diagram that meets specific requirements. One key consideration is the availability of 
sufficient data to train an accurate MLIAP. In order to facilitate the identification and 
correction of any errors in the initial implementation of the workflow, it is also beneficial
to choose a phase diagram that is relatively simple. With these criteria in mind, we chose 
the Cu-Ag-Au ternary system, a choice further supported by the fact that the MLIAPs for this
phase diagram have already been optimized and trained in our previous work. \cite{SNAPbi2ter}.

\begin{figure}[!h]
    \centering
    \includegraphics[width=8cm]{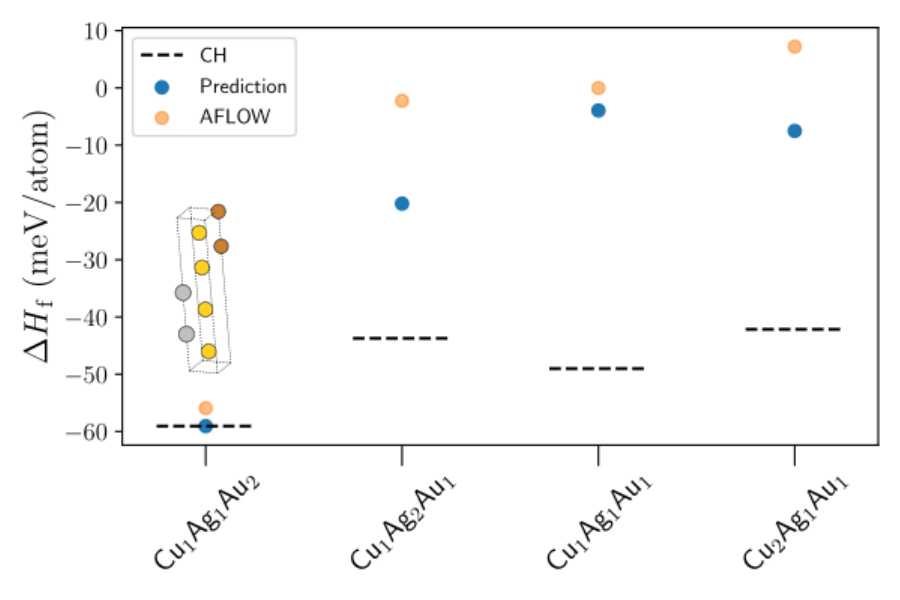}
    \caption{Workflow predictions for the Cu-Ag-Au ternary system across different
    stoichiometries. The graph presents the different compositions and their corresponding
    distance from the convex hull, $\Delta H\mathrm{_{f}}$. The blue points are associated 
    with the predictions from the proposed workflow, whereas the orange ones represent the
    lowest-energy AFLOWlib points. The dashed line (CH) marks the tie-plane position of the convex
    hull. The unit cell of the newly discovered crystal structure on the convex hull is presented 
    as well. Here, Au atoms are in gold, Ag in silver, and Cu in bronze. The proposed workflow 
    manages to identify one stable intermetallic phase among these, namely Cu$_{1}$Ag$_{1}$Au$_{2}$. 
    Furthermore, it manages to outperform the AFLOW dictionary method in all of the presented cases.}
    \label{fig:results_AgAuCu1}
\end{figure}

As a proof of concept, we have focused on the equiatomic Cu$_{1}$Ag$_{1}$Au$_{1}$ ternary phase 
as well as phases with stoichiometric ratios of 2-1-1 and 2-2-1. The reason for this choice is 
that data at these stoichiometries are available in AFLOWlib for comparison. The results of 
the workflow are presented in Figure~\ref{fig:results_AgAuCu1} and Table \ref{ta:results_AgAuCu2}. 
In order to quantitatively assess the stability of the structures proposed by the workflow, 
we use the distance, $\mathrm{\delta}$, from the reference convex hull (AFLOWlib). A negative 
value indicates that the predicted structure lies below the calculated convex hull, establishing 
its stability as an intermetallic compound. Then, the convex hull needs to be recalculated and
corrected by taking into account the newly predicted stable structure. In contrast, a positive distance 
from the convex hull provides a criterion for assessing whether the structure is 
metastable or unstable. In Table~\ref{ta:results_AgAuCu2} values predicted by the workflow 
(AFLOWlib) are labelled as $\mathrm{\delta^{WP}}$ ($\mathrm{\delta^{AFLOW}}$). 

\begin{figure}[!h]
    \centering
    \includegraphics[width=8cm]{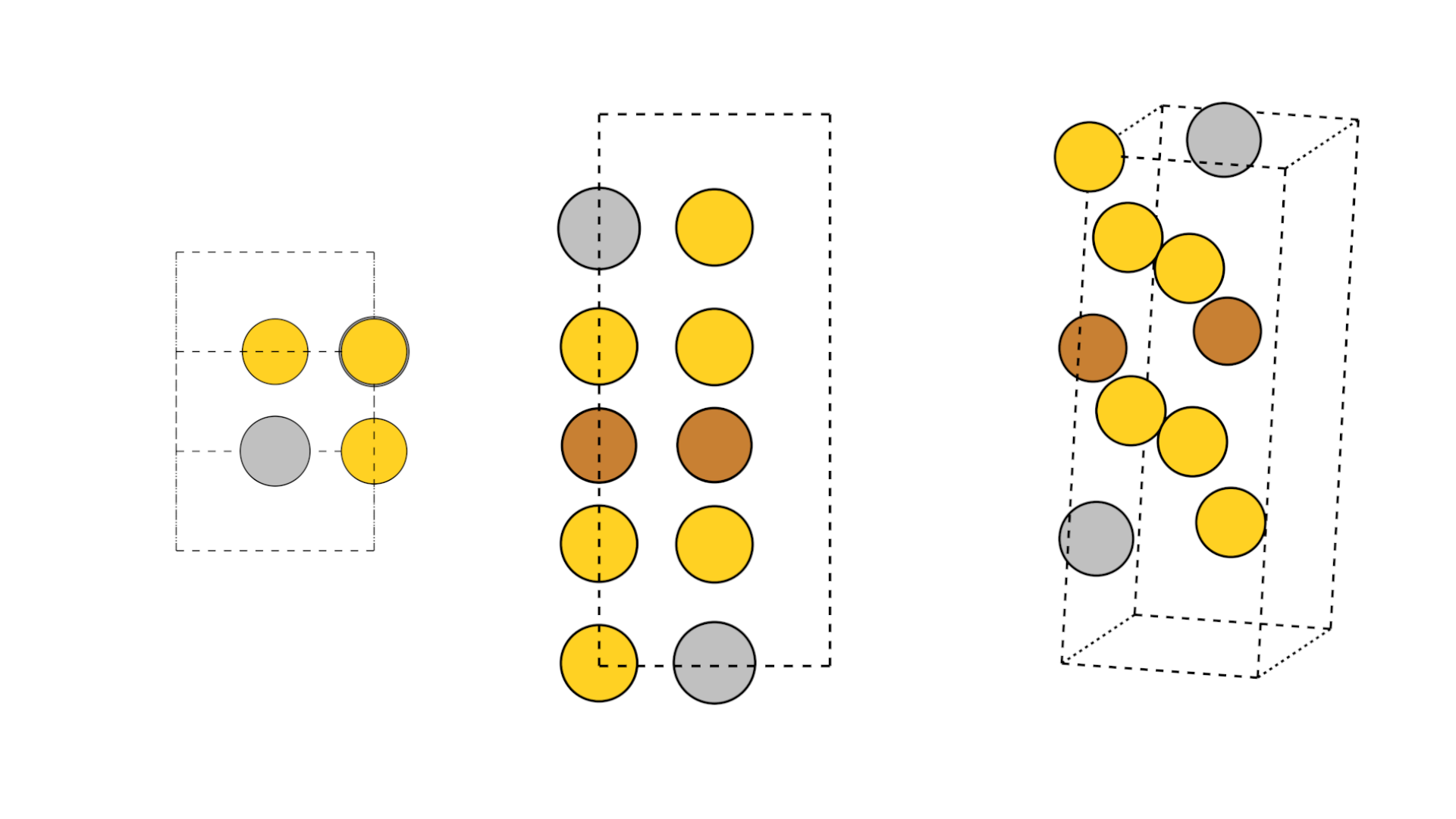}
    \caption{The unit cell of the 1-1-3 crystal structure found on the convex hull, namely 
    Cu$_{1}$Ag$_{1}$Au$_{3}$, is presented in both a top-view with respect to the $z$-axis 
    (left), a side-view along the $x$-axis (middle) and a tilted view (right). In this 
    structure, Au atoms are colored in gold color, Ag atoms in silver color, and copper Cu 
    atoms in bronze.}
    \label{fig:Cu1Ag1Au3_CS}
\end{figure}

The scalability and speed of the algorithm allow us, in principle, to investigate more regions of 
the phase diagram, in a single study, than a pure DFT phase diagram construction scheme. 
This is exemplified by using the proposed workflow to predict structures that are not in 
AFLOWlib's database, namely compounds with 3-1-1 stoichiometry. The results of the benchmark 
are presented in Table~\ref{ta:results_AgAuCu2}, alongside the crystal structure of the new
stable phase, Cu$_{1}$Ag$_{1}$Au$_{3}$, in Figure \ref{fig:Cu1Ag1Au3_CS}.
\begin{table}[!h]
    \centering
    \scalebox{1.10}{\begin{tabular}{c|c|c}
        \hline
         \thead{ Stoichiometry} & \thead{$\mathrm{\delta^{AFLOW}} \; (\mathrm{meV/atom})$} &\thead{$\mathrm{\delta^{WP}} \; (\mathrm{meV/atom})$} \\ 
         \hline
         \hline
         Cu$_{2}$Ag$_{2}$Au$_{1}$& 208.95 &25.99 \\
         Cu$_{2}$Ag$_{1}$Au$_{2}$& 205.69 &37.45 \\
         Cu$_{1}$Ag$_{2}$Au$_{2}$& 90.27  &17.21 \\
         \hline
         \hline
        Cu$_{3}$Ag$_{1}$Au$_{1}$& -- &  20.35\\
        Cu$_{1}$Ag$_{3}$Au$_{1}$& -- &  31.05\\
        Cu$_{1}$Ag$_{1}$Au$_{3}$& -- &  -0.02\\
         \hline
    \end{tabular}
    }
    \caption{Workflow predictions for the Cu-Ag-Au ternary system with 1-2-2 and 1-1-3 
    compositions. The stoichiometries and their corresponding distance from the convex hull,
    $\mathrm{\delta^{WP}}$, are presented. For the 1-2-2 compounds, the distance from the convex 
    hull of the phases available in the AFLOWlib database, $\mathrm{\delta^{AFLOW}}$, are also 
    given. Note that for all materials, the distance from the AFLOWlib convex hull tie plane is 
    used as reference. A new gold-heavy intermetallic, namely Cu$_{1}$Ag$_{1}$Au$_{3}$ is 
    predicted as stable (see Fig.~\ref{fig:Cu1Ag1Au3_CS}).}
    \label{ta:results_AgAuCu2}
\end{table}

Our approach outperforms the AFLOW dictionary method in all cases, demonstrating a better 
predictive capability, which arises from the exploration of a larger pool of prototypes.
Interestingly, the structures predicted by the proposed workflow are consistently closer 
to the convex hull than those predicted by the AFLOW dictionary method. This is to be 
expected, since the workflow effectively selects the relevant structures for creating the 
pool of ternary candidates. Furthermore, our model consistently predicts structures with 
negative or almost negative (< 10 meV/atom) enthalpy of formation, a fact that gives us 
confidence in the reliability of the predicted structures. Notably, we have been able to 
identify two new gold-heavy stable phases, namely Cu$_{1}$Ag$_{1}$Au$_{2}$ and
Cu$_{1}$Ag$_{1}$Au$_{3}$.  This indicates that stable
intermetallic phases may exist on the gold side of the phase diagram. We have confidence 
in our prediction, given the fact that the dictionary method structure for 
Cu$_{1}$Ag$_{1}$Au$_{2}$ is within 3~meV/atom of the convex hull, suggesting the possibility 
of the existence of a stable phase. This is consistent with the formation of the solid
solutions in the gold-rich region of the experimental phase diagram~\cite{prince1990phase}. 
The rest of the structures are considered to be potentially metastable, with an average
distance from the convex hull of around 30~meV/atom~\cite{Metastable}. Overall, our analysis
demonstrates the ability of the workflow introduced here to predict structures closer to 
the convex hull than those from the state-of-the-art dictionary method and possibly 
uncover novel phases, should these exist.

\subsection{Mo-Ta-W ternary convex hull}
\label{subsec:MoTaW}

As a second benchmark, we wish to explore a phase diagram that exhibits a variety of stable 
phases. Thus, the major criterion for our selection, among all the possible transition-metal
ternary combinations, is the total number of stable compounds. The Mo-Ta-W ternary system
emerged as a good candidate, based on a search run with the AFLOW REST API~\cite{restapi}. 
In fact, it exhibits the highest number of stable ternary phases of the entire database of
transition metal alloys. In order to compare our proposed workflow with the dictionary method, 
we have made predictions corresponding to the same stoichiometries presented in the previous
section. Furthermore, we have used our method to explore areas of the phase diagram poorly
covered by AFLOWlib. 

\begin{figure}[!h]
    \centering
    \includegraphics[width=8cm]{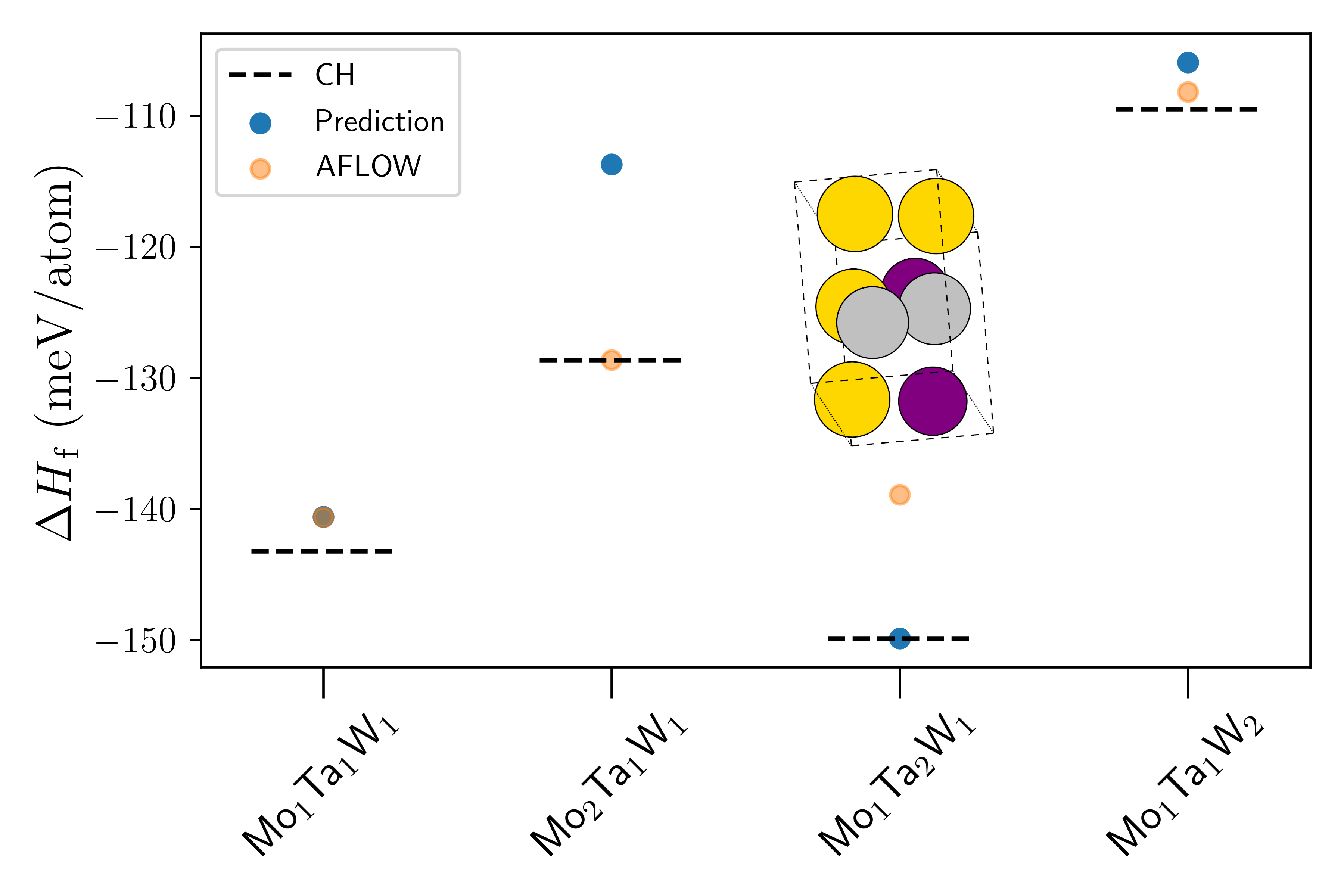}
    \caption{Workflow predictions for the Mo-Ta-W ternary system across different stoichiometries, 1-1-1 and 1-1-2. The graph presents the different compositions and 
    their corresponding distance from the convex hull, $\delta$. The blue 
    points are associated to the predictions from the proposed workflow, whereas the orange
    ones represent the lowest-energy AFLOWlib data. The dashed line marks the tie-plane 
    position of the convex hull (CH). The proposed workflow has managed to identify one previously 
    unknown intermetallic phase, namely Mo$_{1}$Ta$_{2}$W$_{1}$, whose unit cell is shown
    as an inset. Here Mo atoms are in purple, Ta in gold, and W in silver.}
    \label{fig:results_MoTaW1}
\end{figure}

We now perform a similar analysis as that described in the previous section. The structure
prototypes used for the element decoration are extracted from those of the binaries closest 
to their respective convex hulls. Then, an ensemble of ML models relax the created structures and order
them based on their predicted energy. A set of 15 structures for each stoichiometry, 
corresponding to those with the lowest predicted energies, is sampled and proceeds to the 
next stage. This consists in performing a DFT relaxation and a static calculation for each 
one of these predictions. A significant difference with respect to the Cu-Ag-Au system is 
that we now use AFLOWlib's database to train the models without any further re-calculation. 
The AFLOW REST-API is used to download the energies and the crystal structures for the three 
binary convex hulls (Mo-W, Ta-W, and Mo-Ta). The models are trained as explained in the 
Methods section (see Section~\ref{sec:Methods}). Recycling data already available on AFLOWlib 
allows us to avoid about 1,500 DFT relaxation calculations, some of them for cells up to 46 
atoms, just for the training of the model. The results for the 1-1-1 and 1-1-2
compositions, those heavily populated in AFLOWlib, are presented first in  
Figure~\ref{fig:results_MoTaW1}.

In this case as well we predict a new stable intermetallic phase, Mo$_{1}$Ta$_{2}$W$_{1}$. 
However, this time our workflow is not consistently outperforming the dictionary method. 
In fact, for two out of the four stoichiometries investigated in Figure ~\ref{fig:results_MoTaW1},
we obtain compounds with energies similar to the ones already present in AFLOWlib, while for 
one, Mo$_{2}$Ta$_{1}$W$_{1}$, our search delivers a compound with a higher energy. 
Interestingly, in this last case our newly found structure and the original one, contained
in AFLOWlib, belong to different space groups. The AFLOW-predicted one has space group 107
(tetragonal), while our scheme finds a low-stmmetry monoclinic crystal structure with space
group 9. The final geometries are not equivalent as determined by the AFLOW-SYM
tool~\cite{hicks2018aflow}. Nevertheless, the compound discovered with the workflow only 
has an enthalpy of formation 14.91 meV/atom higher than the AFLOWlib compound.

As a force-field approach, our workflow gets better when the MLIAP improves. In this case, 
we have extracted the data used to train the SNAPs from the AFLOWlib repository, a detail 
that led to a less accurate force field than the one used for the Cu-Ag-Au system. In fact, 
minor inconsistencies in the energy data may generate errors in the force-field \cite{deringer2021gaussian,bayerl2022convergence}.
That being understood, we have still demonstrated that new phases can be predicted by an almost 
DFT-free workflow, since our initial data for model training are readily available in the AFLOWlib database. 
The workflow systematically assesses a wide range of compositions and potential compounds. 
Specifically, it involves the evaluation of 331,734 ternaries based on their calculated SNAPs
energies. Following this, the 15 lowest-enthalpy structures, for each stoichiometry, undergo 
relaxation through DFT. Interestingly, the DFT analysis reveals that, on average, the most stable 
compound ranks 7$^\mathrm{th}$ among the suggested options. Additionally, the \textit{ab-initio} 
computations are shortened since all compounds move closer to their equilibrium geometry after the 
SNAP-guided relaxation, in contrast to their fully unrelaxed counterparts.

Perhaps a more accurate force field would also be able to find the AFLOWlib minimum 
for Mo$_{2}$Ta$_{1}$W$_{1}$ (see Fig.~\ref{fig:results_MoTaW1}). Nevertheless, our workflow 
is already able to identify the majority of the structures close to the convex hull. It 
should also be noted that this is the phase diagram for which AFLOWlib's dictionary 
method works best, as it is able to detect four intermetallic phases, more than any other
transition metal alloy phase diagram.

Then, we move to analyse stoichiometries poorly explored in AFLOWlib, namely 1-2-2 and 1-1-4. 
In Table~\ref{ta:results_MoTaW} we provide a comparison of the distance from the convex 
hull for the structures predicted with our method, $\mathrm{\delta}^\mathrm{{WP}}$, and the 
ones from AFLOWlib, $\mathrm{\delta}^\mathrm{{AFLOW}}$. For these compositions, the AFLOWlib
compounds are unstable as they all have a positive enthalpy of formation. In contrast, those
found by our workflow all have a negative enthalpy of formation and are found near or at the
convex hull. These results provide a comparison between our method and AFLOWlib for structures
predicted as unstable by the latter.
\begin{table}[!h]
    \centering
    \scalebox{1.10}{
    \begin{tabular}{c|c|c}
        \hline
         \thead{ Stoichiometry} & \thead{$\mathrm{\delta}^\mathrm{{AFLOW}} \mathrm{(meV/atom)}$} & \thead{$\mathrm{\delta}^\mathrm{{WP}} \mathrm{(meV/atom)}$} \\ 
         \hline
         \hline
         Mo$_2$Ta$_2$W$_1$ & 880.90  & 0.00  \\
         Mo$_1$Ta$_2$W$_2$ & 962.84  & 0.00  \\
         Mo$_2$Ta$_1$W$_2$ & 1032.50 & 8.50  \\
         Mo$_4$Ta$_1$W$_1$ & 320.95  & 46.56 \\
         Mo$_1$Ta$_4$W$_1$ & 516.30  & 3.25 \\
         Mo$_1$Ta$_1$W$_4$ & 334.16  & 0.33 \\
         \hline
    \end{tabular}
    }
    \caption{Workflow predictions for the Mo-Ta-W ternary system with 1-2-2 and 1-1-4 compositions. 
    The stoichiometries and their corresponding distance from the convex hull, 
    $\delta$, are presented ($\mathrm{\delta}^\mathrm{{WP}}$ is for compounds generated by
    our workflow, while $\mathrm{\delta}^\mathrm{{AFLOW}}$ is for the AFLOWlib compounds). 
    Three intermetallic phases are predicted as stable and two others
    metastable. Surprisingly, our algorithm is able to find structures with an energy of up
    to 1~eV/atom lower than those identified by the dictionary method of AFLOWlib.}
    \label{ta:results_MoTaW}
\end{table}

The ability of our workflow to consistently predict structures that are (i) close to the 
convex hull and (ii) have a negative enthalpy of formation is thus demonstrated. The former
point means that we have an effective algorithm to use for structure search in regions 
of interest. The latter validates our physical intuition behind the assumption that the crystal
structures of the binary alloys close to the convex hull can be used as a template for 
atomic decoration in the search for ternary phases. This approach has allowed us to identify
three new intermetallic compounds, see Table~\ref{ta:results_MoTaW}, namely Mo$_{2}$Ta$_{2}$W$_{1}$, Mo$_{1}$Ta$_{2}$W$_{2}$, and Mo$_{1}$Ta$_{1}$W$_{4}$. Such positive
results demonstrate the value of the enhanced freedom in the structure search provided by 
our algorithm with respect to dictionary methods.

\begin{figure}[!h]
    \centering
    \includegraphics[width=8cm]{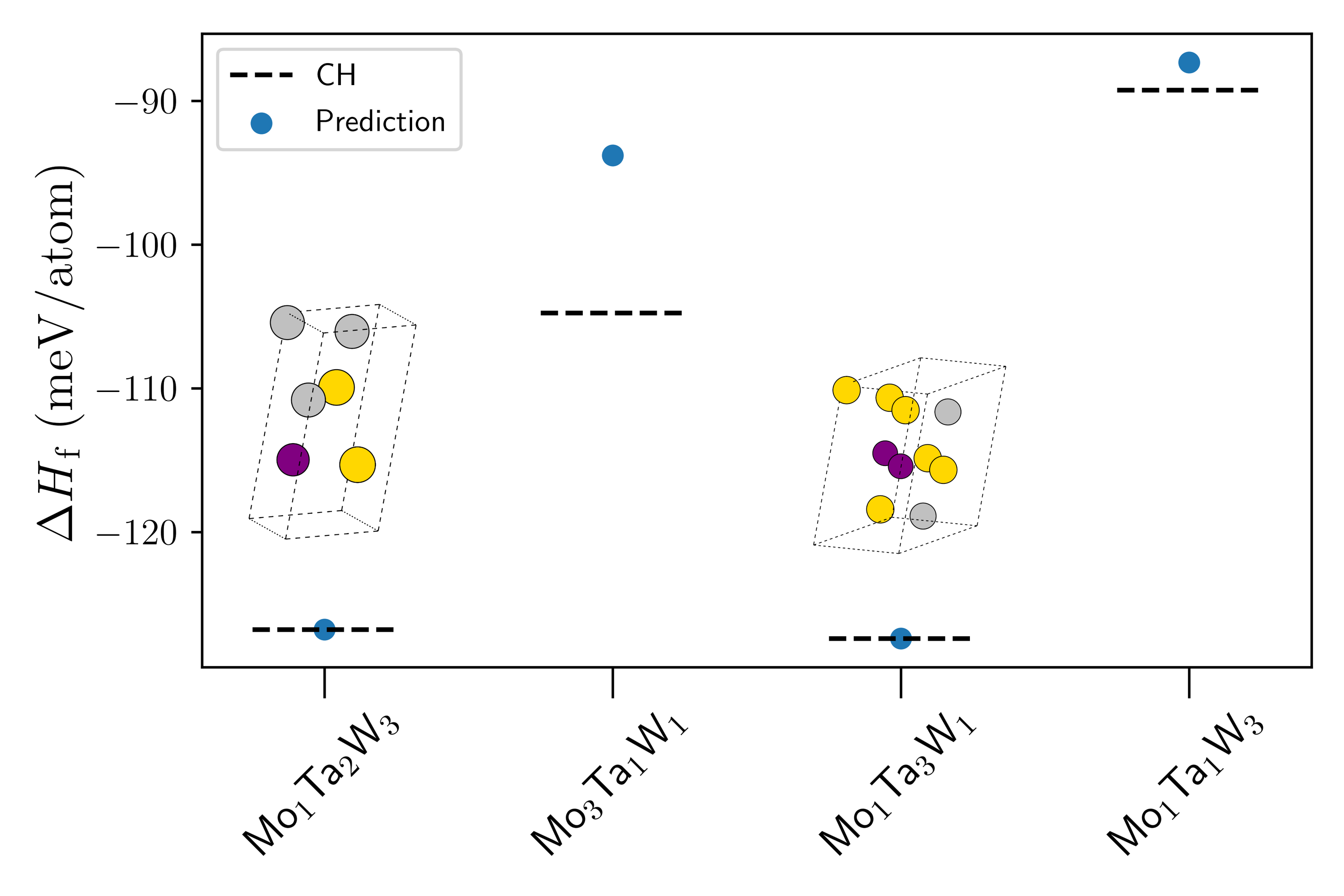}
    \caption{Workflow predictions (blue points) of the enthalpy of formation for the 
    Mo-Ta-W ternary system across the 1-2-3 and 3-1-1 compositions. The enthalpy of 
    formation for each composition at the appropriate the convex hull tie-plane is 
    shown as a dashed line. 
    The unit cell of the crystal structures found on the convex hull are presented as 
    well. Here, Mo atoms are in purple, Ta in gold, and W in silver. Two new
    intermetallics alloys have been identified,  namely
    Mo$_{1}$Ta$_{2}$W$_{3}$ and Mo$_{1}$Ta$_{3}$W$_{1}$.}
    \label{fig:results_MoTaW_new}
\end{figure}

\begin{figure}
    \centering
    \includegraphics[width=8cm]{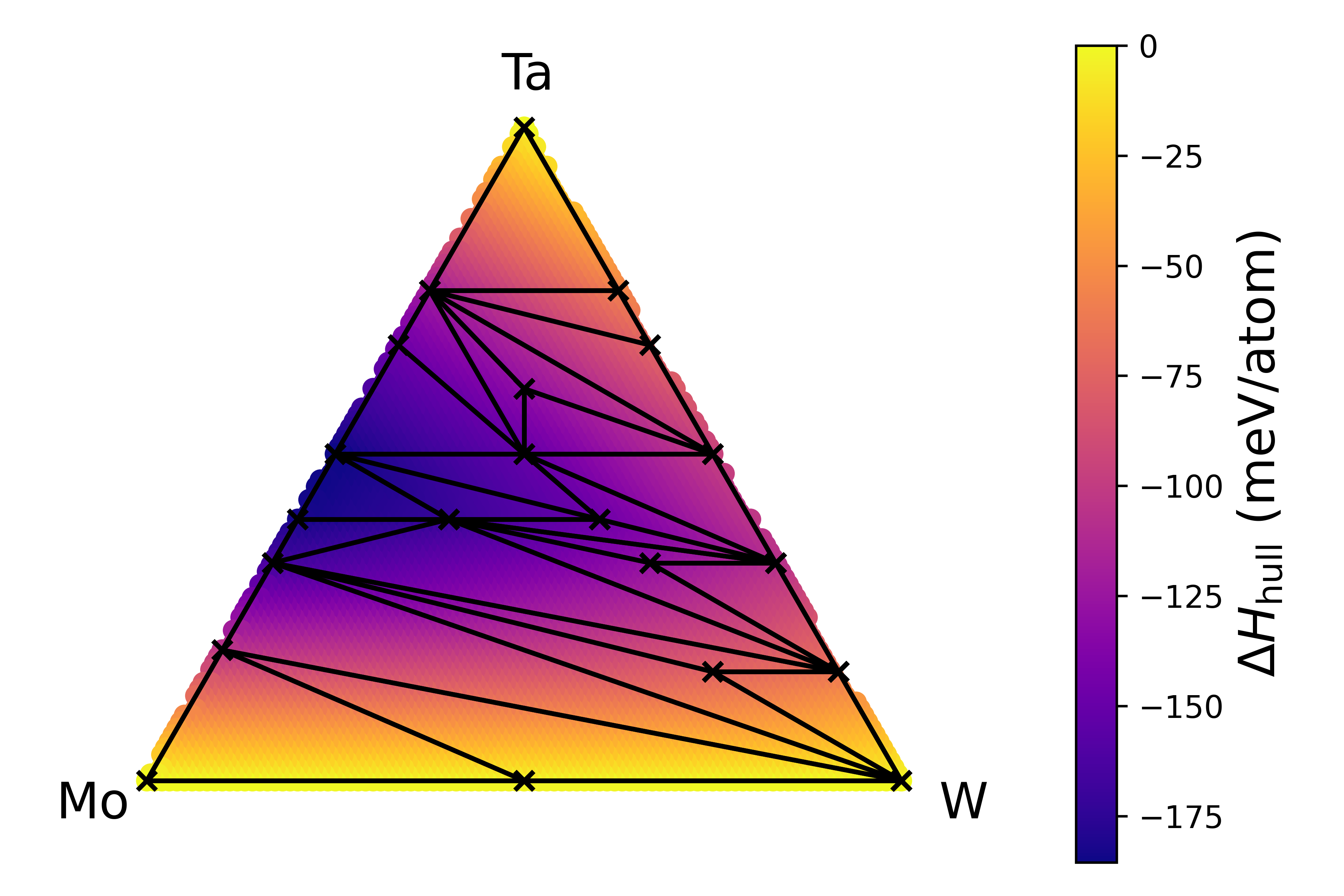}
    \includegraphics[width=7.5cm]{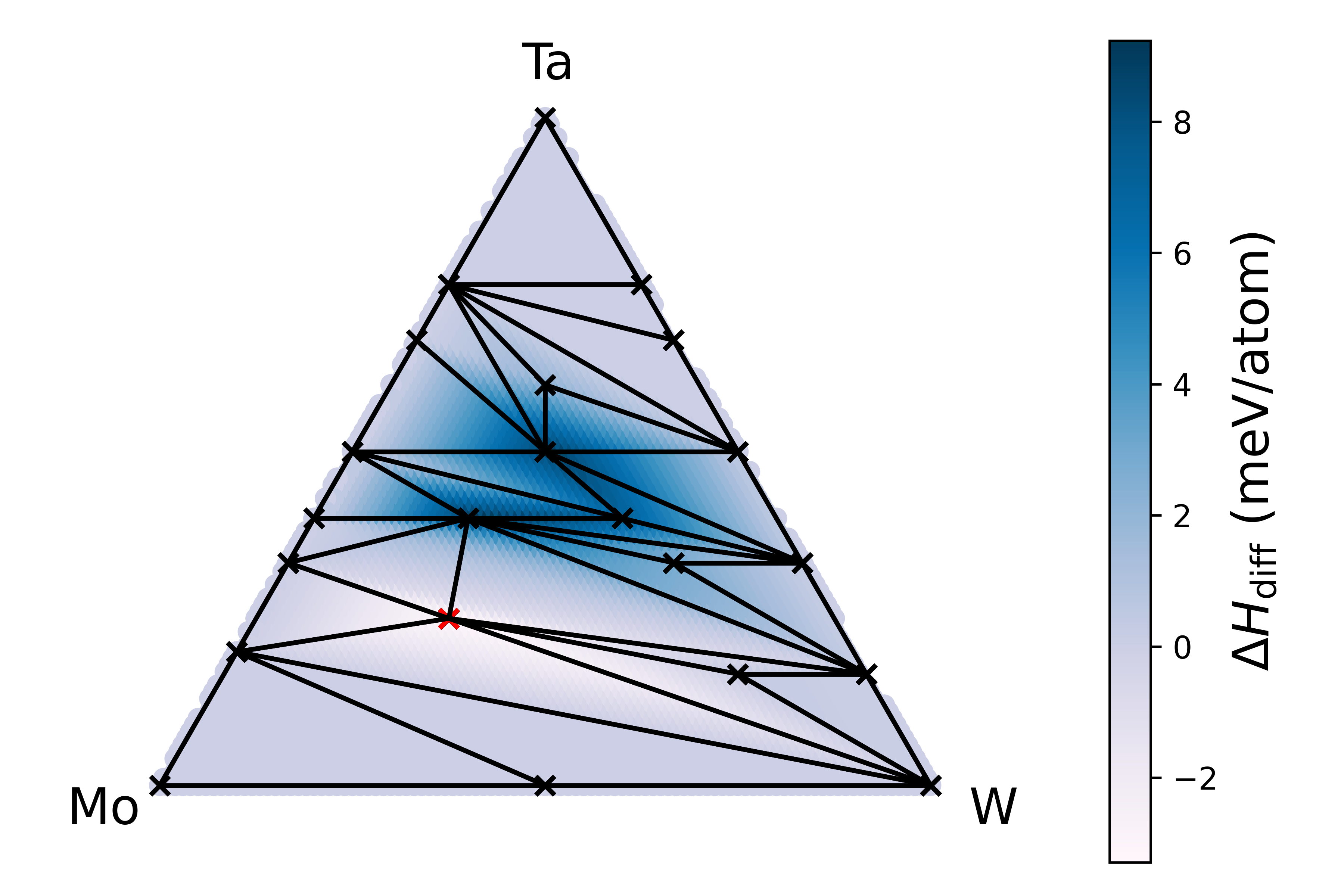}
    \caption{Workflow-computed convex hull for the Mo-Ta-W system (upper panel). The color 
    heat map corresponds to the calculated enthalpy of formation at a given stoichiometry.
    In the lower panel, we present the difference between the convex hull of AFLOWlib 
    (reference) and that computed by our workflow. Black crosses for the ternary region 
    symbolize the newly predicted intermetallic phases, and the red cross denotes the only 
    stable intermetallic originally predicted by the AFLOW dictionary method.}
    \label{fig:results_MoTaW3}
\end{figure}

Finally, following the same spirit as for the analysis of the Cu-Ag-Au system, we now look
into previously unexplored areas of the ternary convex hull. Our results for the 1-2-3 and 1-1-3 compositions are shown in Figure \ref{fig:results_MoTaW_new}.
As one can observe, together with structures away from the tie-plane, we also find
two new stable compounds, namely Mo$_{1}$Ta$_{2}$W$_{3}$ and Mo$_{1}$Ta$_{3}$W$_{1}$. Such
new phases, together with the low-energy ones previously discussed, 
call for a modification of the ternary convex hull that exists in AFLOWlib.
The new diagram is here presented in the top panel of Figure~\ref{fig:results_MoTaW3}.
In order to facilitate the comparison, the lower panel of the same figure shows the difference
between the AFLOW- and our workflow-predicted convex hulls (positive values mean that
our predicted convex hull is lower in energy than the original AFLOWlib one).

The new convex hull returns a picture, where most of the stable ternary structures identified
belong to the Ta-W heavy area, and only one intermetallic alloy exists in the Mo-rich region
of the compositional space. The latter is the compound found on AFLOWlib. Interestingly, the 
new phases predicted by our workflow undercut Mo$_1$Ta$_1$W$_1$, Mo$_1$Ta$_2$W$_1$ and 
Mo$_1$Ta$_1$W$_2$, the other intermetallic alloys initially predicted as stable by AFLOWlib. 
These are now 2.59, 10.95 and 1.31 meV/atom, respectively, above their associated tie-planes 
and have to be considered metastable.
Experimentally, there is evidence that the Mo-Ta-W system forms a ternary solid
solution~\cite{WMoTaphasediagram} at finite temperature across the entire 
phase diagram. It should be noted that the Mo-Ta binary space is far better sampled 
by AFLOWlib than the Ta-W and Mo-W ones. This could imply that it is more difficult to 
reach the convex hull close to such facet of the diagram. In contrast, the Mo-W system only
displays small enthalpies of formation for the stable binary phases, implying that both 
Mo and W form more stable phases with Ta than amongst themselves. These two reasons could
explain why it is more difficult to find stable intermetallic phases in the Mo-rich part 
of the composition space.

\section{Conclusion}
\label{sec:Conclusion}

We have developed a workflow that predicts the crystal structure and assesses the 
stability of ternary compounds of a particular stoichiometry. A library of prototype 
structures is formed from the lowest-enthalpy alloys of the associated binary
subsystems. From this database, derivative ternary structures are generated by site 
decoration. Then, an ensemble of SNAP force-fields is used to select the most 
promising structures among them, bypassing the majority of the \textit{ab-initio} 
calculations. Therefore, the 
proposed workflow highly increases the throughput in the crystal-structure search without
compromising the quality of the predictions. This is used here to map the ternary convex 
hull of transition-metal alloys. The crucial aspect of the proposed scheme is that both 
the training of the force fields and the creation of the prototype ternary structures 
are based solely on the knowledge of the binary phases. As such, no additional DFT 
calculations are required, since both the structures and their corresponding energies 
are readily available on the AFLOWlib database. Employing \textit{ab-initio} 
calculations solely in the final stage of the workflow and focusing them on the most 
promising candidates, allows us to perform a comprehensive exploration of the phase diagram 
of a ternary system with only a few hundred DFT calculations.
This enables us to map previously unexplored portions of the ternary space and to identify regions of interest, thus driving the discovery of novel compounds.

We have demonstrated that the proposed workflow is able to predict crystal structures 
with negative enthalpy of formation and effectively identify the stable intermetallics, 
should they exist. In particular, we have used the Cu-Ag-Au and Mo-Ta-W ternary systems 
as an example. In the first case, we have predicted several new phases that, although not 
all thermodynamically stable, have an enthalpy of formation lower than those found by the 
AFLOW dictionary method. In addition, we have identified an Au-rich composition region, 
where stable intermetallic phases are expected, in accordance with the location of solid 
solutions in the experimental phase diagram~\cite{prince1990phase}. Interestingly, in the 
case of Mo-Ta-W, one of the ternary systems with the largest number of stable intermetallics 
in AFLOWlib, our method is capable of identifying a plethora of new phases, resulting in 
the correction of the original DFT-calculated convex hull proposed by AFLOW.

In summary, we have developed a novel way to integrate machine learning to accelerate a DFT 
workflow. Although the ML model introduced here does not perform as well as force-fields with
tailor-made databases, its construction requires no new DFT calculations and simply recycles
pre-existing results, already present on large-scale databases. This represents an example
of how machine-learning interatomic potentials can be seamlessly integrated into a materials
design pipeline without the need to generate {\it ad hoc} large training sets.

\section{Computational Methods}
\label{sec:CompMethods}

The details of the computational methods are presented in this section. The parameters used for 
the DFT calculations run with VASP~\cite{VASP} are first discussed. A brief presentation of the 
SNAP~\cite{SNAP} is then given, along with details of the implementations used for the current work. 

\subsection{DFT Calculations}
\label{subsec:DFT}

All DFT calculations are performed using the Vienna Ab initio Simulation 
Package (VASP)~\cite{VASP}, version 5.4.4. Projector augmented wave (PAW) pseudopotentials 
are used for each element together with the Perdew-Burke-Ernzerhof (PBE) functional
\cite{PhysRevLett.77.3865}. A plane wave cutoff of 600~eV is used for all calculations. The 
energy convergence criterion for each self-consistent cycle is of $10^{-4}$~eV. Full atomic 
relaxations are performed (update of atomic positions, cell volume and lattice parameters) 
with a stopping criterion on the forces of $10^{-3}$ eV/\AA. A Fermi-Dirac smearing of 0.2~eV 
is chosen for all calculations.

For the $k$-point sampling, a Gamma-centered mesh is employed for all calculations. The density 
of the mesh and the spacing between $k$-points is chosen based on AFLOWlib's convergence criteria
\cite{10.1016/j.commatsci.2015.07.019}. The mesh is system specific and determined from the 
$N_\mathrm{KPPRA}$ (number of $k$-points per reciprocal atom). The number of sampling points 
along each direction is proportional to the norm of the corresponding reciprocal lattice vector. 
The total number of sampling points per reciprocal atom is then minimised and $N_\mathrm{KPPRA}$ 
is used as a lower bound. Values of $10\times 10^{3}$ and $6\times 10^{3}$ are used for static 
calculations and relaxations respectively. 

\subsection{Spectral Analysis Neighbor Potential}
\label{subsec:SNAP}

The Spectral Analysis Neighbor Potential (SNAP)~\cite{SNAP} is used as an energy predictor. As 
described in section~\ref{subsec:EnsembleSNAP}, an ensemble of models is employed for predictions. 
Equation \eqref{eq:SNAPE} defines the expression of the function, $E_\mathrm{SNAP}$, and combined 
with equation \eqref{eq:MLIAPE}, gives the energy of a system with $N$ atoms. The atomic fingerprints 
that define the chemical environments of each atom $i$ in the system, belonging to species $\alpha_{i}$, 
are the bispectrum components~\cite{bartok2013representing}. These are used to represent configurations 
instead of seemingly more obvious choices (e.g. atomic Cartesian coordinates), as they are invariant 
upon rotation and permutations of identical atoms. Note that invariance with respect to translations is guaranteed
by Eq.~\eqref{eq:MLIAPE}. For each atom, the vector $\mathbf{B}_{i}^{\alpha_{i}}$, which collects the first 
components up to a maximum index, is taken as a feature for the machine-learning model (ridge regression 
in the case of SNAP). A short description of the bispectrum components is given below. 

The neighborhood of an atom $i$ atom can be described by a density function, $\rho_{i}$, centered at 
that atom with delta functions at the sites of surrounding atoms, within a sphere of radius $r_\mathrm{cut}$. 
It is defined in three dimensions as
\begin{equation}
    \rho_{i}\left ( \mathbf{r} \right ) = \delta \left ( \mathbf{r} - \mathbf{r}_{i} \right ) + \sum_{j} w^{\alpha_{j}} \delta \left ( \mathbf{r} - \mathbf{r}_{j} \right ) f_{c}\left ( r_{ij} \right ),
\label{eq:density}
\end{equation}
where the sum is over all atoms within $r_\mathrm{cut}$ from the central atom. Here, $\mathbf{r}_{i}$ is 
the position of atom $i$, $r_{ij}=|\mathbf{r}_{i}-\mathbf{r}_{j}|$, $w^{\alpha_{j}}$ is the specie-specific 
weight of atom $j$ and $f_{c}$ is a cut-off function that smoothly runs to zero as $r_{ij}$ approaches
$r_\mathrm{cut}$, as defined in \cite{NNP}. In order to represent this density distribution as a vector, 
it is expanded in a suitable basis. Atomic positions are first mapped onto the 4D sphere, by switching to 
polar coordinates $\left ( \theta ,\phi ,r \right )$ and by defining a third polar angle, $\theta_{0}$, 
from the radial coordinate (see Ref.~\cite{bartok2013representing} for details). The density function is 
then expanded in terms of hyperspherical harmonics $U^{J}_{m',m}$, the natural basis for expansion on 
the 4D sphere. Dropping the atomic index, $\rho$ is written as
\begin{equation}
\rho\left ( \mathbf{r} \right ) = \sum_{J=0}^{\infty }\sum_{m,m'=-J}^{J}c_{m',m}^{J}U^{J}_{m',m}\left ( \theta ,\phi ,\theta _{0} \right )\:.
\label{eq:expansion}
\end{equation}
The hyperspherical harmonic index $J$ runs in half-integer steps, while $m$ and $m'$ run between 
$-J$ and $J$ in integer steps. The outer sum is truncated in practice at a value $J_\mathrm{max}$, 
treated as a hyperparameter. The expansion coefficients, $c_{m,m'}^{J}$, cannot be used as descriptors, 
since they are complex and are not invariant under system rotation. From them, however, the 
rotationally-invariant and real-valued bispectrum components $B_{J,J_{1},J_{2}}$ are constructed
\begin{align*}
B_{J,J_{1},J_{2}} &= \sum_{m_{1}',m_{1}=-J_{1}}^{J_{1}}c_{m'_{1},m_{1}}^{J_{1}}\sum_{m_{2}',m_{2}=-J_{2}}^{J_{2}}c_{m'_{2},m_{2}}^{J_{2}}\\
&\quad \times \sum_{m',m=-J}^{J} C_{mm_{1}m_{2}}^{J,J_{1},J_{2}}C_{m'm'_{1}m'_{2}}^{J,J_{1},J_{2}}\left ( c_{m',m}^{J} \right )^{*}\:.
\label{eq:bispectrum}
\end{align*}
Here, $C_{mm_{1}m_{2}}^{J,J_{1},J_{2}}$ and $C_{m'm'_{1}m'_{2}}^{J,J_{1},J_{2}}$ are the Clebsch-Gordan 
coefficients, which possess the same symmetry invariances as the system. After taking the non-zero and 
unique distinct components, the bispectrum vector is formed, denoted $\mathbf{B}_{i}^{\alpha_{i}}$ with 
atomic and specie indices. The bispectrum components are a highly non-linear representation of the local 
atomic coordinates and account for up to four-body interactions. Their complexity is what makes it possible 
for them to be effectively used together with a simple regressor in SNAP to accurately map structures to 
energies. 

The fitting, testing and predictions of the SNAP models used are performed using an in-house python library 
built with \textsc{scikit-learn} \cite{scikit-learn} and the Atomic Simulation Environment (\textsc{ase} 
\cite{ASE}) python libraries. The bispectrum components are computed using \textsc{lammps} \cite{LAMMPS}. 
The pipeline is built in python to perform the API download of binary structures and energies from the 
AFLOWlib \cite{AFLOW-CHULL} database and to generate derivative structures from the prototypes 
using \textsc{enumlib} \cite{hart2008}. DFT calculations are managed by using a combination of 
\textsc{ase} \cite{ASE} and \textsc{pymatgen} \cite{pymatgen}.

\begin{acknowledgements}

This work has been supported by the Irish Research Council Advanced Laureate Award (IRCLA/2019/127),
and by the Irish Research Council postgraduate program (MC). We acknowledge the DJEI/DES/SFI/HEA Irish 
Centre for High-End Computing (ICHEC) and Trinity Centre for High Performance Computing (TCHPC) for the 
provision of computational resources. 

\end{acknowledgements}

\section*{Author Contributions}

This section is written according to the CRediT system.
H.R. and M.M. contributed equally to this work. H.R. and M.M. contributed to conceptualisation, methodology, software, data curation, formal analysis, investigation, validation, visualisation, writing the original draft, as well as reviewing and editing the manuscript. M.C. contributed to conceptualisation, software as well as reviewing and editing the manuscript. S.S. contributed to conceptualisation, funding acquisition, project administration, resources, supervision, as well as reviewing and editing the manuscript.


\bibliography{tex.bib}
\bibliographystyle{ieeetr}

\end{document}